\documentclass{JHEP3}

\usepackage{graphicx}
\newcommand{\be}{\begin{equation}}
\newcommand{\ee}{\end{equation}}
\newcommand{\ben}{\begin{eqnarray}}
\newcommand{\een}{\end{eqnarray}}
\def\pls{\partial\!\!\!/}
\def\bs{b\!\!\!/}
\def\ps{p\!\!\!/}
\def\As{A\!\!\!/}

\def\m{\mu}

\title{A Remark on Lorentz Violation at Finite Temperature}

\author{Tiago Mariz, Jose R. Nascimento, Eduardo Passos and Rubens F.
Ribeiro\\
Departamento de F\'\i sica, Universidade Federal da Para\'\i ba,
Caixa Postal 5008, 58051-970 Jo\~ao Pessoa,
Para\'\i ba, Brazil\\
E-mail: {tiago, jroberto, passos, rfreire@fisica.ufpb.br}}
\author{Francisco A. Brito\\
Departamento de F\'\i sica, Universidade Federal de Campina Grande,
58109-970 Campina Grande, Para\'\i ba, Brazil\\
E-mail: {fabrito@df.ufcg.edu.br}}

\abstract{We investigate the radiatively induced Chern-Simons-like
term in four-dimensional field theory at finite temperature. The
Chern-Simons-like term is temperature dependent and breaks the
Lorentz and CPT symmetries. We find that this term remains
undetermined although it can be found unambiguously in different
regularization schemes at finite temperature.\\

\vspace{1cm}

}

\maketitle

\begin{document}

\section{Introduction}

In the last years, it has been discussed in the literature if the
Lorentz and CPT symmetries are exact or not
\cite{Ja,Go,D,CG,D1,CO,JAK,MP,CH,ags,brett03,brett04}. Theoretical
investigations have pointed out that these symmetries can be
approximate. The spontaneous breakdown of Lorentz symmetry was first
considered in electrodynamics \cite {bjorken} and string theory
\cite{alan}. The modern quantum field theory also admits such
phenomenon from the theoretical point of view. In other words, the
basic laws of nature have Lorentz and CPT symmetries, but the vacuum
solution of the theory could spontaneous violate these symmetries.
This mechanism is identical to Higgs mechanism for particle field
theory.

The standard model extension \cite{D1} which preserves underlying
properties that are renormalizability, unitarity and gauge
invariance, in the vacuum state appears to be filled up by fields
that select a fixed direction in spacetime, and explicitly violate
Lorentz and CPT symmetries. The realization of this violation can be
obtained in the QED by adding the term $ \frac{1}{2}
k_{\mu}\epsilon^{\mu\alpha\beta\gamma}F_{\alpha\beta}A_{\gamma} $ to
the Maxwell's theory --- this is a Chern-Simons-like term with a
constant vector $k_{\mu}$ in four dimensions --- and another term
which is a CPT-odd term for fermions, i.e., $\bar\psi b_{\mu}
\gamma^{\mu} \psi $ with a constant vector $b_{\mu}$ \cite{Ja}. Such
an extension of QED does not break the gauge symmetry of the action
and equations of motion but it does modify the dispersion relations
for different polarization of photons and Dirac's spinors.

The dynamical origin of the parameters $k_{\mu}$ and $b_{\mu}$
present in the Lorentz and CPT symmetry breaking has called
attention of a great number of people and is an interesting problem
to be analyzed. In this respect, a relation between $k_{\mu}$ and
$b_{\mu}$ is obtained when we integrate over the fermion fields in
the modified Dirac action such that radiative corrections may lead
to $k_{\mu}=C b_{\mu}$. This result introduces a modification of the
Electrodynamics, which allows for the explicit violation of Lorentz
and CPT symmetries. The issue has been carefully investigated in
several different contexts leading to results where $C$ does vanish
\cite{CG} and results where $C$ does not \cite{JAK}. See
Refs.~\cite{ja99,GB,chen,mpv}, for further details on such issues.
In this work we will analyze the behavior of the parameter $C$ when
we take temperature into account. We do this by using derivative
expansion method of the fermion determinant \cite{Fr,sh,zuk,LHC,das}
and the imaginary time formalism. By comparing with results in the
literature \cite{znr,griguolo,ebert}, we find that although this
term can be found unambiguously in different regularization schemes
at finite temperature, it remains undetermined. We then conclude
that its value can be only fixed by phenomenological constraints
\cite{mpv,brtlm00,brtlm05}.

\section{The Model}

In this section we shall induce the Chern-Simons-like term
\cite{CRH}. CPT and Lorentz symmetries are violated in the fermionic
sector as \be \label{lag} {\cal L} = \bar\psi\left[i\pls - m -
\gamma_5\bs - e\As\right]\psi, \ee where $b_{\mu}$ is a constant
4-vector which selects a fixed direction in the space-time.

The corresponding generating functional is \be \label{fg} Z[A]=\int
D\bar\psi(x)D\psi(x)\exp\left[i\int{\cal L}\;\;d^4x\right] \ee We
substitute Eq.(\ref{lag}) into Eq.(\ref{fg}) and integrate over the
fermion field  to obtain \be \label{Seff} Z[b,A] = {\rm Det}(i\pls -
m - \gamma_5\bs - e\As)= \exp[iS_{eff}[b,A]]. \ee Since we assume
the field $A_{\mu}$ as an external field, no Legendre transformation
is required to go from the connected vacuum functional to effective
action. Then, the effective action takes the form \be \label{se}
S_{eff}[b,A] = -i\,{\rm Tr}\ln [i\pls - m - \gamma_5\bs - e\As]. \ee
Here the symbol Tr stands for the trace over Dirac matrices, trace
over the internal space as well as for the integrations in momentum
and coordinate spaces. Note that the equation (\ref{se}) above can
be written as \be S_{eff}[b,A] = S_{eff}^{(0)}[b] +
S_{eff}^{(1)}[b,A], \ee where
\ben &&S_{eff}^{(0)}[b]=-i\,{\rm Tr}\ln[i\pls-m-\gamma_5\bs],\\
\label{snew} &&S_{eff}^{(1)}[b,A]=i\int_0^1{dz\,{\rm
Tr}\left[\frac{1}{i\pls - m - \gamma_5\bs -
z\,e\As(x)}\,e\As(x)\right]}. \een Since the term $S_{eff}^{(0)}[b]$
is independent of the gauge field and cannot induce Chern-Simons
term, we shall focus only on the second term $S_{eff}^{(1)}[b,A]$.
To perform the momentum space integration in (\ref{snew}) it is
convenient to consider the prescription \cite{LHC}: \be
i\,\pls\to\ps,\qquad\qquad
\As(x)\to\As\left(x-i\frac{\partial}{\partial p}\right).\ee The
equation (\ref{snew}) now reads\ben \label{seff1} S_{eff}^{(1)}[b,A]
&=& i \int_0^1 dz \int{d^4x}\int\frac{d^4p}{(2\pi)^4}\,{\rm
tr}\left[\frac{1}{\ps - m - \gamma_5\bs -
ze\As(x-i\frac{\partial}{\partial p})}e\As(x)\right]. \een Let us
make use of the relation
$$
\frac{1}{A-B} = \frac{1}{A} + \frac{1}{A}B \frac{1}{A} +
\frac{1}{A}B \frac{1}{A}B \frac{1}{A} + \cdots
$$
and consider $A = \ps - m-\gamma_5\bs$ and $B=
ze\As(x-i\frac{\partial}{\partial p_\m})$. We can manipulate the
Eq.(\ref{seff1}) to keep only first order derivative terms which are
linear in $\bs$ and quadratic in $\As$. Carrying out the integral in
$z$ gives \ben \label{seff2} S_{eff}^{(1)}[b,A] &=&
-i\frac{e^2}{2}\int d^4 x \int \frac{d^4 p}{(2 \pi)^4}
\nonumber \\
 &\times& {\rm tr} \left[\frac{1}{\ps - m}i\partial_{\m}\As
 \frac{\partial}{\partial p_{\m}}\frac{1}{\ps - m}\gamma_5
 \bs\frac{1}{\ps - m}\right.\As \nonumber \\
&+& \frac{1}{\ps - m}\gamma_5\bs \frac{1}{\ps - m}i
\left.\partial_{\m}\As\frac{\partial}{\partial p_{\m}}\frac{1}{\ps
- m}\As \right], \een where we have used the relation
$$
\frac{\partial}{\partial p_\mu}\frac{1}{\ps -m} = -\frac{1}{\ps -m}
\gamma^{\mu}\frac{1}{\ps-m}.
$$
Now taking the traces of the products of $\gamma$ matrices on
relevant terms, i.e., the terms that contain
$tr\,\gamma_5\gamma^{\mu}\gamma^{\nu}\gamma^{\alpha}\gamma^{\beta}$,
the  Eq.(\ref{seff2}) takes the form \be \label{seff3}
S_{eff}^{(1)}[b,A] = -\frac{e^2}{2}\int d^4\, x\int \frac{d^4\,
p}{(2 \pi)^4} \frac{N}{(p^2 - m^2)^4}, \ee where $N$ is given by
\ben \label{numer} N &=& -4i(p^2 -
m^2)\left[\epsilon^{\alpha\beta\m\sigma}\left(3m^2+p^2\right)-4
\epsilon^{\alpha\beta\m\nu} p_{\nu} p^{\sigma}\right] \\ \nonumber
&\times&b_\sigma
\partial_{\mu}A_\alpha A_\beta.
\een Note that by power counting the momentum integral in
Eq.~(\ref{seff3}) have terms with logarithmic divergence. Let us use
the relation \cite{MP,JAK,mpv}\be \label{re}
\int\frac{d^Dq}{(2\pi)^D}q_{\mu}q_{\nu}f(q^2)=\frac{g_{\mu\nu}}{D}
\int\frac{d^Dq}{(2\pi)^D}q^2f(q^2), \ee that naturally removes the
logarithmic divergence. Now, considering $D=4$, the terms containing
$p^2$ and $p_{\nu}p^{\sigma}$ in (\ref{numer}) cancel out and we
find \be N = -12m^2i(p^2 - m^2)\epsilon^{\alpha\beta\m\sigma}
b_\sigma
\partial_{\mu}A_\alpha A_\beta.
\ee  In this way, the logarithmic divergence in (\ref{seff3})
disappears, so that the effective action now reads \ben
\label{seff33} S_{eff}^{(1)}[b,A]& =&\left[ {6im^2 e^2}\int
\frac{d^4\, p}{(2 \pi)^4} \frac{1}{(p^2 - m^2)^3}\,\right] \\
\nonumber & \times & \epsilon^{\alpha\beta\m\sigma}{b_{\sigma}}\int
d^4\,x
\partial_{\mu}A_\alpha A_\beta,
\een which is finite by power counting. Evaluating the  momentum
integral in the (\ref{seff33}) we obtain unambiguously the
Chern-Simons coefficient \cite{JAK} \be \label{k} k_{\mu} =
\frac{3e^2}{16\pi^2}b_{\mu}. \ee However, if we use another
regularization scheme $k_{\mu}$ may vanish, as for instance, in
Pauli-Villars regularization scheme \cite{D1}. The fact that the
value of $k_{\mu}$ depends on the regularization scheme, corresponds
to an ``ambiguity'', i.e., finite values but undetermined ones
\cite{ja99}. This issue has been well discussed in the literature
\cite{ja99,GB,chen,mpv}. Next we study such undetermined coefficient
when we take temperature into account.

\begin{figure}[h]
\centerline{\includegraphics[{angle=90,height=7.0cm,angle=180,width=8.0cm}]
{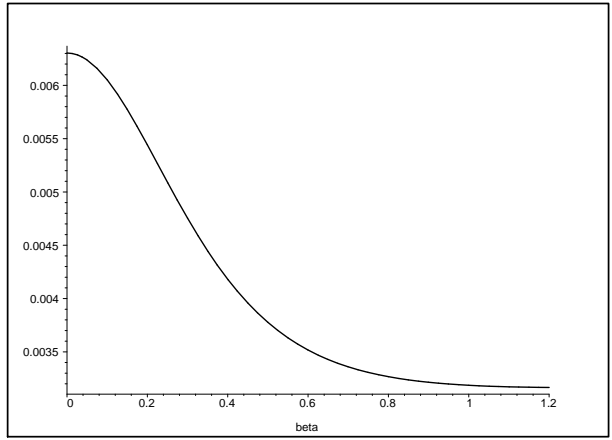}} \caption{The function $f(m^2, \beta)$ is diferent from
zero everywhere. At zero temperature ($\beta\to\infty$), the
function tends to a nonzero value ${1}/{32\pi^2}$.}\label{fig2}
\end{figure}

%%%%%%%%%%%%%%%%%%%%%%%%%%%%%%%%%%%%%%%%%%%%%%%%%%%
%%%%%%%%%%%%%%%%%%%%%%%%%%%%%%%%%%%%%%%%%%%%%%%%%%%
%%%%%%%%%%%%%%%%%%
Let us now assume that the system is at thermal equilibrium with a
temperature $T=1/{\beta}$. In this case we can use Matsubara
formalism, for fermions, which consists in taking $p_0 = (n +
1/2)2\pi/{\beta}$ and changing $(1/2\pi)\int dp_0=1/\beta\sum_n$
\cite{DJ}. We also change the Minkowski space to Euclidean space, by
making $x_0 = -ix_4$, $p_0=ip_4$ and $b_0=ib_4$, such that
$p^2=-p_E^2$, $p_E^2 = {\bf p}^2 + p_4^2$, $d^4p=id^4p_E$  and
$d^4x=-id^4x_E$. Now the Eq.(\ref{seff33}) can be written as \ben
\label{seff5} S_{eff}^{(1)}[b,A] = 6e^2 f(m^2, \beta)
\epsilon^{\alpha\beta\m\sigma}{b_{\sigma}} \int(-i)d^4\,x_E
\partial_{\mu}A_\alpha A_\beta,
\een where $f(m^2, \beta)$, is the Chern-Simons coefficient
dependent on the temperature which is given by \ben \label{f} f(m^2,
\beta)&=&\frac{m^2}{\beta}\int\;\frac{d^3{\bf p}
}{(2\pi)^3}\sum_{n=-\infty}^{\infty}
\frac{1}{({\bf p}^2+p_4^2 + m^2)^3}\nonumber \\
&=&\frac{m^2}{2\beta}\frac{d^2}{d(m^2)^2}\int\;\frac{d^3{\bf p}
}{(2\pi)^3}\sum_{n=-\infty}^{\infty}\frac{1}{({\bf p}^2+p_4^2 +
m^2)^2}. \een We calculate the momentum integral by adopting
dimensional regularization scheme to obtain \be \label{t} f(m^2,
\beta) = \frac{m^2}{2\beta}\frac{\Gamma(3-D/2)}{(4\pi)^{D/2}}
\sum_{n=-\infty}^{\infty} \frac{1}{(p_4^2 + m^2)^{3-D/2}}. \ee To
perform summation we shall use below an explicit representation for
the sum over the Matsubara frequencies \cite{FO}: \ben\label{fo}
\sum_n [(n+b)^2 + a^2]^{-\lambda}&=& \frac{\sqrt{\pi}\Gamma(\lambda
- 1/2)}{\Gamma(\lambda)(a^2)^{\lambda - 1/2}}
\\  &+&\nonumber
4\sin(\pi\lambda)\int_{|a|}^\infty \frac{dz}{(z^2 - a^2)^{\lambda}}
Re\left(\frac{1}{\exp 2\pi(z + ib) -1}\right), \een which is valid
for $1/2<\lambda<1$. This implies that for $\lambda=3-D/2$ as given
in Eq.(\ref{t}) we cannot apply this relation for $D=3$, because the
integral in (\ref{fo}) does not converge. Thus, let us perform the
analytical continuation of this relation, so that we obtain \ben
&&\int_{|a|}^\infty \frac{dz}{(z^2 - a^2)^{\lambda}}
Re\left(\frac{1}{\exp 2\pi(z + ib) -1}\right)= \nonumber \\
\nonumber &=& \frac{1}{2a^2}\frac{3-2\lambda}{1-\lambda}
\int_{|a|}^\infty \frac{dz}{(z^2 - a^2)^{\lambda-1}}
Re\left(\frac{1}{\exp 2\pi(z + ib) -1}\right)\\  &-&
\frac{1}{4a^2}\frac{1}{(2-\lambda)(1-\lambda)} \int_{|a|}^\infty
\frac{dz}{(z^2 - a^2)^{\lambda-2}}
\frac{d^2}{dz^2}Re\left(\frac{1}{\exp 2\pi(z + ib) -1}\right). \een
Now for $D=3$ the Eq.(\ref{t}) takes the form \be f(m^2, \beta) =
\frac{1}{32\pi^2}+\frac{1}{16}F(\xi), \ee where $\xi=\frac{\beta
m}{2\pi}$ and the function \be F(\xi)=
\int_{|\xi|}^{\infty}dz(z^2-\xi^2)^{1/2} \frac{\tanh(\pi
z)}{\cosh^2(\pi z)}, \ee approaches the limits:
$F(\xi\to\infty)\to0$ ($T\to0$) and $F(\xi\to0)\to{1}/{2\pi^2}$
($T\to\infty$) --- see Fig.\ref{fig2}. Thus, we see that at high
temperature the Chern-Simons coefficient is twice its value at zero
temperature, i.e., $f(m^2,\beta\to0)={1}/{16\pi^2}$. On the other
hand, at zero temperature, one recovers the result (\ref{k}).
%%%%%%%%%%%%%%%%%%%%%%%%%%%%%%%%%%%%%%%%%%%%%%%%%%%%%%%%%%%%%%%%%%%%
%%%%%%%%%%%%%%%%%%%%%%%%%%%%%%%%%%%%%%%%%%%%%%%%%%%

\section{Conclusions}

We have studied the induction of Chern-Simons-like term at finite
temperature. We adopted dimensional regularization to evaluate
momentum integrals. Our result is finite, but does not fully agree
with other results in the literature. We argue that this is due to
different regularization schemes.  We find that at zero temperature
limit we recover the result found at zero temperature in
Refs.~\cite{JAK,MP}. In such limit our result leads to a nonzero
Chern-Simons-like term, a behavior also predicted in
Ref.\cite{znr}-obtained with the use of dimensional regularization-
and the result in Ref.\cite{ebert}- obtained with the use of cut off
regularization scheme. However, it is in conflict with the result
found in Ref.~\cite{griguolo} which suggests the vanishing of the
Chern-Simons-like term at zero temperature. On the other hand, at
high temperature our result behaves as the result of
Ref.\cite{griguolo}. But now, however, it conflicts with the results
in Ref.\cite{znr} and in Ref.\cite{ebert} which predict that the
Chern-Simons-like term vanishes at high temperature. These results
are all finite, and they show that the Chern-Simons-like coefficient
is indeed undetermined just as it happens at zero temperature ---
see Ref.\cite{ja99} for discussions on such issues. Our result
supports the understanding that the Chern-Simons-like term can only
be determined by phenomenological constraints, as already emphasized
in Ref.~\cite{mpv}.

\acknowledgments The authors would like to thank D. Bazeia for
discussions and  CAPES, CNPq, PROCAD/CAPES, PRONEX/CNPq/FAPESQ and CNPq/FINEP/PADCT
 for partial support.

\end{document}